\documentclass[final]{svjour2}
\usepackage{graphicx}
\usepackage{rotating}
\usepackage{amssymb}
\usepackage{mathptmx}
\usepackage[numbers]{natbib}
\makeatletter
\journalname{Journal of Low Temperature Physics}
\bibpunct{}{}{,}{s}{}{,}

\begin{document}

\newcommand{\hdblarrow}{H\makebox[0.9ex][l]{$\downdownarrows$}-}
\title{Zel'dovich-Starobinsky Effect in Atomic Bose-Einstein Condensates: Analogy to Kerr Black Hole}

\author{Hiromitsu Takeuchi$^1$ \and Makoto Tsubota$^1$ \and Grigory E. Volovik$^{2,3}$}

\institute{1: Department of Physics, Osaka City University, Sumiyoshi-ku, Osaka 558-8585, Japan\\
Tel.: 011+81-06-6605-2501\\
Fax: 011+81-06-6605-2522\\
\email{hiromitu@sci.osaka-cu.ac.jp}\\
2: Low Temperature Laboratory, Helsinki University of Technology, P. O. Box 2200, FIN-02015 HUT, Finland\\
3: L. D. Landau Institute for Theoretical Physics, Moscow 119334, Russia\\ }

\date{08.10.2007}

\maketitle

\keywords{Zel'dovich-Starobinsky Effect, atomic Bose-Einstein condensate, Kerr black hole }

\begin{abstract}
 We consider circular motion of a heavy object in an atomic Bose-Einstein condensate (BEC) at $T=0{\rm K}$. Even if the linear velocity of the object is smaller than the Landau critical velocity, the object  may radiate quasiparticles and thus experience the quantum friction. The radiation process is similar to Zel'dovich-Starobinskii (ZS) effect --  the radiation by a rotating black hole. This analogy emerges when one introduces the effective acoustic metric for quasiparticles. In the rotating frame this metric has an ergosurface, which is similar to the ergosurface  in the metric of a rotating black hole. In a finite size BEC, the quasiparticle creation takes place when the ergosurface is within the condensate and occurs via quantum tunneling from the object into the ergoregion. The dependence of the radiation rate on the position of the ergosurface is investigated.

PACS numbers:
 03.75.Nt,
%Other Bose-Einstein condensation phenomena
 03.75.Kk
%Dynamic properties of condensates; collective and hydrodynamic excitations, superfluid flow
04.62.+v
%Quantum field theory in curved spacetime
04.70.Dy
%Quantum aspects of black holes, evaporation, thermodynamics
\end{abstract}
%%%%%%%%%%%%%%%%%%%%%%%%%%%%%%%%%%%%%%%%%%%%%%%%%%%%%%%%
\section{Introduction}
%%%%%%%%%%%%%%%%%%%%%%%%%%%%%%%%%%%%%%%%%%%%%%%%%%%%%%%%
 There are many challenges to simulate the phenomena on black holes by using condensed matter systems \cite{artificial}.
 One of them is to study an analogy between gravity and superfluidity, in which a superfluid ground state (superfluid vacuum) serves as the analog of the vacuum of  relativistic quantum fields, while the flow of the superfluid liquid imitates the metric field acting on `relativistic' quasiparticles (phonons, ripplons, fermionic excitations, etc.) \cite{volovik}.
 The radiation from the black holes can be tested in this analogy, for example the Hawking radiation \cite{artificial}. In this work we discuss how using  atomic Bose-Einstein condensates one can simulate the Zel'dovich-Starobinskii (ZS) effect \cite{zeld,staro} --  the amplification and spontaneous emission of electromagnetic waves by a body rotating in the quantum vacuum, or by a rotating black hole (Kerr-BH). Radiation of electromagnetic mode occurs if its frequency $\omega$ and its angular  momentum quantum number $l$ satisfy the condition  $\omega - l\Omega <0$, where $\Omega$ is the angular velocity of the rotating body. This corresponds to the negative energy of the mode in the frame co-rotating with the body. In the Kerr black hole the negative energy states exist in the ergoregion,
where  the speed of the frame dragging exceeds that of light. In the ergoregion $g_{00}<0$ \cite{Landau}.
 The boundary of the ergoregion $g_{00}=0$, where the speed of frame dragging is equal to the speed of light, is called the ergosurface, 
 %%%%%%%%%%%%%%%%%%%%%%%%%%%%%%%%%%%%%%%%%%%%%%%%%%%%%%%%%%%%%

\section{The Zel'dovich-Starobinskii effect in superfluid vacuum}
%%%%%%%%%%%%%%%%%%%%%%%%%%%%%%%%%%%%%%%%%%%%%%%%%%%%%%%%%%%%%%%%%%
Phonon excitation in a superfluid moving with velocity ${\bf v}_s$ has the Doppler shifted energy $\epsilon ({\bf p})=cp+{\bf v}_s \cdot {\bf p}$. From this dispersion relation it follows that phonon trajectory in moving liquid obeys the null-geodesic equation $ds=0$, where  $ds$ is the interval determined by the acoustic metric introduced by Unruh\cite{Unruh}
\begin{equation}
ds^2=g_{\mu\nu}dx^\mu dx^\nu=c^2 dt^2-(d{\bf r}-{\bf v}_sdt)^2~.
\label{metric}
\end{equation}
 In superfluids the speed of light and the frame dragging velocity are played by sound velocity $c$ and superfluid velocity ${\bf v}_s$ correspondingly.  Then in the ergoregion the superfluid velocity satisfies the condition $v_s^2>c^2$, while
 the ergosurface, on which $g_{00}=0$ for the acoustic metric, is defined as a surface with $v_s^2=c^2$.

 Now let us consider  an small object moving in a stationary  superfluid along the circular trajectory with radius $\rho=\rho_{\rm obj}$ and with linear velocity $v_{\rm obj}<c$. Since in BEC the  Landau critical velocity for creation of quasiparticles is $v_{\rm Landau}=c$, naive consideration suggests that the object is not radiating.  However, radiation occurs due to Zel'dovich-Starobinskii effect. To see that let us choose the reference frame rotating with the angular velocity $\Omega= v_{\rm obj}/\rho_{\rm obj}$. In this frame the object is stationary, while the superfluid moves with velocity field  ${\bf v}_s=-\Omega \rho \hat{\phi}$, where we use  the cylindrical coordinate ${\bf r} =(\rho,\phi,z)$. At $\rho>c/\Omega$, the superfluid velocity exceeds the speed of sound and $g_{00}<0$, which demonstrates that $\rho_{\rm erg}=c/\Omega$ marks the position of the ergorsurface. In the ergoregion, i.e. at $\rho>c/\Omega$, the energy of some phonons becomes negative. These are phonons that satisfy  condition  $\omega - l\Omega <0$:
 \begin{eqnarray}
\epsilon ({\bf p})=cp + {\bf v}_s \cdot {\bf p} = cp - \rho p_{\phi} \Omega =\hbar(\omega- l \Omega) < 0~,
\label{eq:ZSinSF}
\end{eqnarray}
It is energetically favorable to create these phonon modes in the ergoregion. However, the source of the radiation  -- the moving body -- is away from the ergoregion: $\rho_{\rm obj} = v_{\rm obj}/\Omega <\rho_{\rm erg}=c/\Omega$. In the process  of radiation the phonon mode created by the source at $\rho=\rho_{\rm obj} $ is tunneling through the barrier of the positive energy region $ \rho_{\rm erg}>\rho >\rho_{\rm obj}$ to the ergoregion  at $\rho > \rho_{\rm erg}$ \cite{Calogeracos}.
In the infinite system the radiation takes place for arbitrary $\Omega$, and thus for arbitrary velocity of the circulating object, though for small $v_{\rm obj}\ll c$ the  ergoregion is far from the object and the probability of quantum tunneling is suppressed by factor   $(\Omega\rho_{\rm obj}/c)^{2l}=(v_{\rm obj}/c)^{2l}$.

%%%%%%%%%%%%%%%%%%%%%%%%%%%%%%%%%%%%%%%%%%%%%%%%%%
\section{The Zel'dovich-Starobinskii effect in atomic Bose-Einstein condensates}
\begin{figure}[hbpt] \centering
  \includegraphics[width=1. \linewidth]{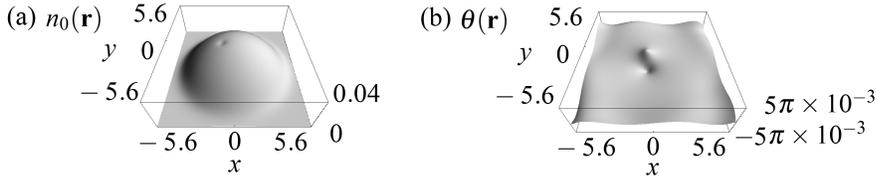}
  \caption{The plots of the density $n_0({\bf r})$ (a) and phase $\theta ({\bf r})$ (b) of the condensate with the object circulating with the angular velocity $\Omega > \Omega_c$.
 (a) The object potential makes a density dip.
 (b) The phase is almost uniform: its  variation is on the order of  $10^{-2}$.
 Here the length $x$ and $y$ are scaled with $b=\sqrt{\hbar/m \omega_{\rm t}}$; and the density $n_0$ of BEC is normalized: $\int n_0 d{\bf r}=1$. 
 }
\label{fig:frictionless_state}
\end{figure}

%%%%%%%%%%%%%%%%%%%%%%%%%%%%%%%%%%%%%%%%%%%%%%%%%%
Now let us consider how the  ZS effect is modified in a finite system -- in an atomic BEC trapped by an axisymmetrical potential at $T=0{\rm K}$. 
The condensate is initially at rest in the laboratory frame. The moving heavy body is reproduced by the additional potential which makes a small dip in the condensate at $\rho=\rho_{\rm obj}$ (see Fig.\ref{fig:frictionless_state} {\it  left}) and this dip is circulating with linear velocity $v_{\rm obj}<c$.  
As in  case of the moving body the circulating dip serves as an environment which provides us with the preferred reference frame.  This is the  frame  rotating with frequency $\Omega=v_{\rm obj}/\rho_{\rm obj}$ in which the trapping potential and the dip are time-independent. 
The main difference from the infinite system discussed above is that in the finite system the 
negative energy states in the rotating frame appear only after $\Omega$  reaches some critical value
$\Omega_c$. As a result the radiation occurs only after the velocity of the object reaches  $v_c= \Omega_c\rho_{\rm obj}$. Thus in the finite system the critical velocity of the circulating object is non-zero, but it is much smaller  than the Landau critical velocity $c$ if the object is close to the center of the condensate.

%%%%%{fig:frictionless_state}%%%%%%%%%%%%%%%%%%%%%%%%%%%%%%%%%%%%%%%%%%%%%%%%%%%%%
%%%%%%%%%%%%%%%%%%%%%%%%%%%%%%%%%%%%%%%%%%%%%%%%%%%%%%%%%%%%%%

\begin{figure}[bpt] \centering
  \includegraphics[width=1. \linewidth]{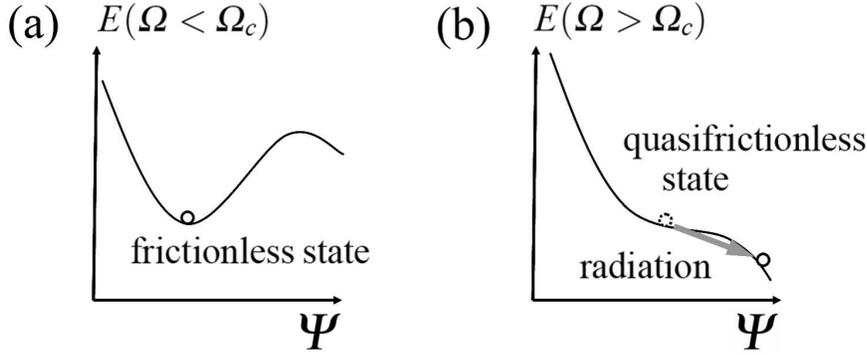}
  \caption{The schematic diagram of the energy of the BEC in the  frame co-rotating with the circulating body.
 (a) The state at $\Omega<\Omega_c$  corresponds to the local minimum and keeps the frictionless circulation of the object. The energy gap which stabilizes the frictionless state is reduced as $\Omega$ is increased.
 (b) The  state at $\Omega>\Omega_c$ is energetically unstable and is destroyed  by the spontaneous radiation of quasiparticles with negative energies in the rotating frame.} 
\label{fig:phase_space}
\end{figure}

The states with  $\Omega<\Omega_c$ are frictionless: they correspond to the local minima of the   energy $E -L_z\Omega$, where $L_z$ is the angular momentum of the superfluid  (Fig.\ref{fig:phase_space}-(a)).  In these states, all the quasiparticles have positive energy in the rotating frame; $\hbar(\omega- l\Omega) >0$.  At $\Omega > \Omega_c$ the stability is lost, the local minimum transforms to the saddle point  (Fig.\ref{fig:phase_space}-(b)), and the moving object would radiate quasiparticles with 
 negative energy $\hbar(\omega- l\Omega) <0$.  

%%%%%{fig:phase_space}%%%%%%%%%%%%%%%%%%%%%%%%%%%%%%%%%%%%%%%%%%%%%%%%
%%%%%%%%%%%%%%%%%%%%%%%%%%%%%%%%%%%%%%%%%%%%%%%%%%%%%%%%%%%%%%%%%%%%%%%%%%%%%%%

Let us illustrate this using  numerical simulations of the BEC.
In the absence of quasiparticles, BEC at $T=0$ can be described by the macroscopic wave function $\Phi({\bf r})=\sqrt{n_0({\bf r})} \exp[i\theta ({\bf r})]$, which satisfies  the Gross-Pitaevskii (GP) equation
\begin{eqnarray}
\mu \Phi= \biggl( -\frac{\hbar^2}{2m} \nabla^2+V_{\rm trap}+V_{\rm obj}+g|\Phi|^2  -\Omega \hat{L}_z \biggr) \Phi~.
\label{eq:GP}
\end{eqnarray}
Here $\hat{L}_z=-i\hbar({\bf r} \times {\bf \nabla})_z$; $g=4 \pi \hbar^2 a/m$; $m$ and $a$ are the atomic mass and $s$-wave scattering length, $V_{\rm trap}$ and $V_{\rm obj}$ refer to the trapping and the object potential, respectively.
 We shall consider the two-dimensional system with a harmonic potential $V_{\rm trap}=1/2m\omega_{\rm t} ^2 \rho^2$, and the object potential $V_{\rm obj}$ is assumed to be localized and repulsive. The frictionless states below $\Omega_c$ are obtained using the imaginary time propagation (ITP) of the Gross-Pitaevskii energy. The states above $\Omega_c$ are also obtained with the ITP if the time of development of instability is long enough.
In Fig.\ref{fig:frictionless_state} the state is shown with $\Omega=0.56 \times \omega_{\rm t}>\Omega_c$, $g=200 \times \hbar \omega b^2/N_{2D}$, where $N_{2D}$ is the particle number per unit length along the $z$ axis and $b=\sqrt{\hbar/m \omega_{\rm t}}$.
 The object potential has a Gaussian form $V_{\rm obj}=V_0 \exp [-({\bf r}-{\bf r}_{\rm obj})^2/w^2] $, where $V_0=0.05 \times m \omega_{\rm t}^2 R_T^2$, ${\bf r}_{\rm obj}=(-0.1 \times R_T,0)$, $w=0.05 \times R_T$ and $R_T=3.99 \times b$ is the Thomas-Fermi radius of the condensate.
 
 Since the object potential is weak, the circulating object makes a small density dip at ${\bf r}={\bf r}_{\rm obj}$ (Fig.\ref{fig:frictionless_state}-(a)), which practically does not disturb the superfluid velocity $(\hbar/m){\bf \nabla} \theta({\bf r})$: it is almost zero everywhere in the laboratory frame (Fig.\ref{fig:frictionless_state}-(b)). The  role of the dip is to provide the perturbation which violates the cylindrical symmetry and induces the nonzero matrix element for the transition between the state without quasiparticles and the states with quasiparticles; the dip also provides the distinguished reference frame which in the absence of other interactions serves as the frame of the environment.
 
%%%%%%%%%%%%%%%%%%%%%%%%%%%%%%%%%%%%%%%%%%%%%%%%%%
\section{Critical velocity of circulating body}
%%%%%%%%%%%%%%%%%%%%%%%%%%%%%%%%%%%%%%%%%%%%%%%%%%

 Let us now find the critical velocity $\Omega_c$, at which the negative energy phonon mode appears in the condensate and thus the  radiation becomes energetically possible. The spectrum of quasiparticles are obtained by coupling the GP eq.(\ref{eq:GP}) and the Bogoliubov de Gennes (BdG) equation.
 For the calculation of the spectrum we may neglect the small deformation of the condensate caused by the object potential  and consider the axisymmetric BEC $\Phi ({\bf r})=\sqrt{n_0(\rho)}$.
 Then the Bose operator is written as $\hat{\Psi}({\bf r})=\Phi (\rho)+\sum'_{k,l}[u_{kl} \hat{\alpha}_{kl}-{v_{kl}}^*{\hat{\alpha}_{kl}}^{\dagger}]$, where $u_{kl}({\bf r},t)=u_{kl}(\rho) \exp[i(l \phi-\epsilon_{kl}t/\hbar)]$ and $v_{kl}({\bf r},t)=v_{kl}(\rho) \exp[i(l \phi-\epsilon_{kl}t/\hbar)]$.
 The numbers $k$ and $l$ refer to the radial and angular quantum number of  quasiparticles and the sum $\sum'$ is taken without a trivial zero-energy mode $(k,l)=(0,0)$.
 As a result the BdG eq. is reduced to \cite{isoshima}
\begin{eqnarray}
\epsilon_{kl}(\Omega) \left[ 
\begin{array}{ccc}
u_{kl}(\rho) \\
v_{kl}(\rho) \\
\end{array} 
\right]
=\left[ 
\begin{array}{ccc}
T_+ & -U \\
U^* & -T_- \\
\end{array} 
\right] \left[ 
\begin{array}{ccc}
u_{kl}(\rho) \\
v_{kl}(\rho) \\
\end{array} 
\right],
\label{eq:BdGcyl}
\end{eqnarray}
where $T_{\pm}=-\hbar^2/2m [(d/d\rho)^2+(1/\rho) (d/d\rho)-l^2/\rho^2]+V_{\rm trap}+2g|\Phi|^2-\mu \mp \hbar l \Omega$ and $U=g\Phi^2$.
 The normalized condition of $u_{kl}$ and $v_{kl}$ is $\int d{\bf r}(|u_{kl}|^2-|v_{kl}|^2)=1$.
 It can be shown that the values $u_{kl}$ and $v_{kl}$ are independent of $\Omega$ but the energy spectrum depends as $\epsilon_{kl}(\Omega)=\epsilon_{kl}(0)-\hbar l \Omega$.

%%%%%{fig:BdG_graph}%%%%%%%%%%%%%%%%%%%%%%%%%%%%%%%%%%%%%%%%%%%%%%%%%%%%%%
\begin{figure}[hbpt] \centering
  \includegraphics[width=1. \linewidth]{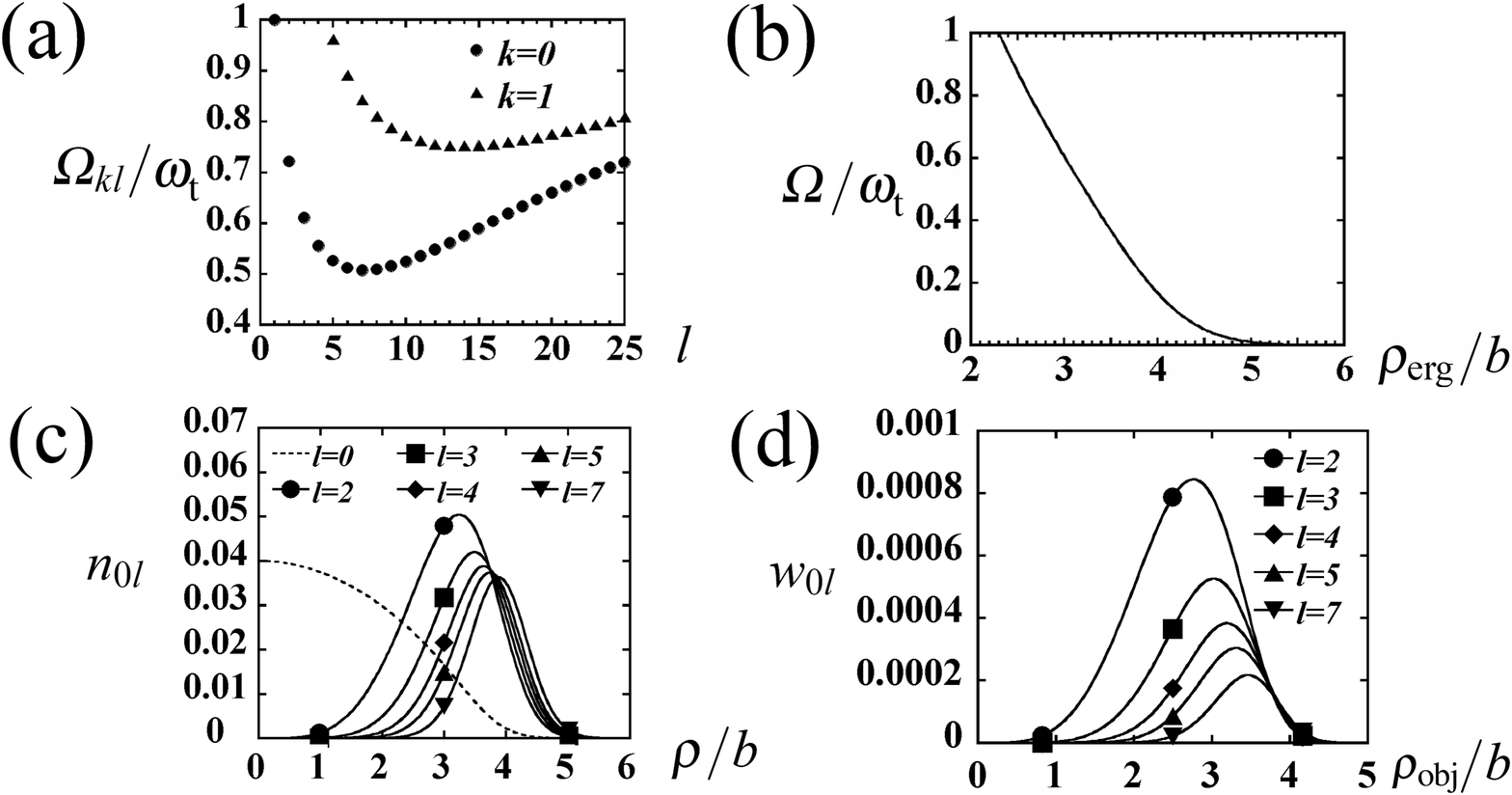}
  \caption{(a) Critical velocity  $\Omega_{kl}$ for nucleation of  modes $(k=0,l)$   and $(k=1,l)$
  as a function of the orbital quantum number $l$. The critical velocity  $\Omega_c$ is given by  the lowest  of them: $\Omega_c=\Omega_7 \simeq 0.508 \omega_{\rm t}$. (b) Position of the ergosurface  $\rho_{\rm erg}$ in a semiclassical description as a function of $\Omega$. (c) Probability density $n_{0l}$ of the modes $(0,l)$; $n_{00} = n_0$ refers to the condensate density. (d) The radiation rate of the mode $l$ by the body is proportional to $w_{0l}(\rho_{\rm obj})=n_0(\rho_{\rm obj})n_{0l}(\rho_{\rm obj})$ at the position of the body. $w_{0l}$  is plotted as a function of the position of the body. 
 }
\label{fig:BdG_graph}
\end{figure}
%%%%%%%%%%%%%%%%%%%%%%%%%%%%%%%%%%%%%%%%%%%%%%%%%%%%%%%%%%%%%%%

 The mode $(k,l)$ can  be radiated if the rotation velocity $\Omega$ reaches the value $\Omega_{kl}$, which satisfies condition $\epsilon_{kl}(\Omega_{kl}) =0$.
 This value  $\Omega_{kl}$ is plotted in Fig.\ref{fig:BdG_graph}-(a) for $g=200 \times \hbar \omega_{\rm t} b^2/N_{2D}$.
 Modes with $(0,l)$ are most important, since modes with higher $k$ have higher $\Omega_{kl}$.
 From Fig. \ref{fig:BdG_graph}-(a)  it follows that the mode $(k,l)=(0,7)$ has the lowest value of the critical velocity; it thus determines  the global critical velocity  at which radiation becomes possible:  $\Omega_c = \Omega_{07} \simeq 0.508 \omega_{\rm t}$.

\section{Radiation by circulating body}

Since quantum numbers of relevant modes  are small, the classical notion of the ergoregion is not well defined. Nevertheless,  ergosurface may be  crudely introduced as the surface $\rho_{\rm erg}=c(\rho_{\rm erg})/\Omega$, where $c(\rho) =\sqrt{g n_0(\rho)/m}$ is the local speed of sound (Fig.\ref{fig:BdG_graph}-(b)). The position of the ergosurface for $\Omega=\Omega_c$ is $\rho_{\rm erg}(\Omega_c) \simeq 3.2$, i.e. at the periphery of the condensate and far from the circulating object. This confirms the general picture that an object circulating  with the subsonic  velocity spontaneously radiates sound modes which are formed far from the object.

This is also supported by exact quantum mechanical consideration. If the circulating object is small  the matrix element of transition is concentrated locally at the position of the object ${\bf r}={\bf r}_{\rm obj}$, and  the radiation rate of the mode $(k,l)$ is proportional to the condensate density $n_0({\bf r}_{\rm obj})$ at the position of the object and to the probability density $n_{kl}({\bf r}_{\rm obj})$ of the mode also at the position of the object, and thus it is determined by the product  $w_{kl}(\rho_{\rm obj})=n_0(\rho_{\rm obj})n_{kl}(\rho_{\rm obj})$.

 The probability density of the $(0,l)$ mode $n_{0l}(\rho)=|u_{0l}(\rho)|^2+|v_{0l}(\rho)|^2$ is shown in Fig.\ref{fig:BdG_graph}-(c). The `penetration' of the mode  into the central region of the condensate is monotonically decreasing as $\rho$ is reduced. In terms of the semiclassical description, this reduction can be interpreted as the  tunneling  of quasiparticles into the energetically forbidden region \cite{volovik}.  The quantity  $w_{kl}(\rho_{\rm obj})=n_0(\rho_{\rm obj})n_{kl}(\rho_{\rm obj})$ which determines the radiation rate   is shown in   Fig.\ref{fig:BdG_graph}-(d). It demonstrates  that the radiation  is enhanced as the object becomes closer to the ergosurface.

%%%%%%%%%%%%%%%%%%%%%%%%%%%%%%%%%%%%%%%%%%%%%%%%%%
\section{Conclusions}
%%%%%%%%%%%%%%%%%%%%%%%%%%%%%%%%%%%%%%%%%%%%%%%%%%
Spontaneous radiation is possible by an external object circulating in BEC with linear velocity much smaller than the Landau critical velocity. The origin of radiation is similar to the ZS radiation from the Kerr black hole. It can be described in terms of  quantum tunneling through the energetically forbidden region between the object and the ergoregion. The radiation rate depends on  rotation speed and on the distance between the object and the ergosurface.  The role of the object might be played by the moving optical potential made by the blue or red detuned laser.  

\begin{acknowledgements}
We would like to thank Ralf. Sch$\ddot{\rm u}$tzhold for useful discussion.
 HT acknowledges the support of a Grant-in-Aid for JSPS Fellows (Grant No. 199748).
 MT acknowledges the support of a Grant-in Aid for Scientific Research from JSPS (Grant No. 18340109) and a Grant-in-Aid for Scientific Research on Priority Areas from MEXT (Grant No. 17071008).
\end{acknowledgements}

\end{document}